# Signatures of quantum coherence in the optical line shape of an exciton in the presence of dynamic disorder

**Rajesh Dutta[1], Kaushik Bagchi[2] and Biman Bagchi[1,*]**

**[1]SSCU, Indian Institute of Science, Bangalore 560012, India,**

**[2]Chitrakut Annexe, Malleswaram, Bangalore 560055, India.**

***Email: profbiman@gmail.com**

## *Abstract*

**We address the effects of quantum coherences on the optical line shape of an exciton in the presence of dynamic disorder. We consider a one dimensional excitonic system that consists of two levels placed at regular intervals. Detailed *analytical calculations* of line shape have been carried out by using Kubo's stochastic Liouville equation (K-QSLE). We make use of the observation that in the site representation, the Hamiltonian of our system with constant off-diagonal coupling *J* is a tridiagonal Toeplitz matrix (TDTM) whose eigenvalues and eigen functions are known analytically. This identification is particularly useful for long chains where the eigen values of TDTM help to understand crossover between static and fast modulation limits. We summarize the new results as follows. (i) In the slow modulation limit when the bath correlation time is large, the effects of spatial correlation are not negligible. Here the line shape is broadened and the number of peak increases beyond the ones obtained from TDTM (constant off-diagonal coupling element J and no fluctuation). (ii) However, in the fast modulation limit when the bath correlation time is small, the spatial correlation is less important (iii) Importantly, we find that the line shape can capture that quantum coherence affects in the two limits differently.**

# I. INTRODUCTION

In recent years, there have been an enormous growth of interest in energy transfer processes both in photosynthesis[1,2] and in conjugated polymers.[3-7] In addition to these important systems, the study of excitonic energy transfer in molecular crystals has a long history. Experimentally one studies not only energy transfer dynamics but also optical line shapes in these systems. Line shape is one of the most important tools by which we study dynamics in correlated systems and thereby understand the role of the different factors in a given energy transfer process. However, most of the studies of coherence have been limited to molecular systems, like studies of electronic and vibrational coherence in molecules, and dimers. Several recent studies have been devoted to the efficiency of energy transfer in photosynthetic and model dimer systems.[8-15] In a recent work Moix, Khasin and Cao[16] have studied the effect of static disorder and dephasing in diffusion of exciton and line shape and concluded that with increasing the dephasing rate, line shape become Lorentzian (as may be expected). However, in the opposite limit i.e. in the limit of small dephasig rate, the line shape is Gaussian. In an extended correlated system, coherence is propagated and promoted by the presence of a significant time independent coupling between different spatially separated groups or sites. Here coherence takes a somewhat different shape than what we encounter in energy relaxation in an isolated molecular system where coherence is discussed among say, different vibrational energy levels.[17-21] The situation is even more complicated in an extended quantum system in presence of dynamical disorder.[22] As far we are aware, there is yet no systematic study of these two factors on the coherence and line shape even on two level dimeric systems. Most of the studies employ quantum master equation formalism which has certain limitations. First, one needs to employ perturbative expansion and truncation (mostly after 2$^{nd}$ order) which is not appropriate for photosynthetic systems. Second, approach is often Markovian. Third, it is difficult to take into account spatial correlations which could be important in a correlated system. To mimic these types of systems and to avoid such kind of approximation one may require a theory which is more rigorous.

There are several lacunae even in the established studies. A connection between population transfer dynamics that easily reflects the presence of coherences through oscillatory time dependence and the optically measured line shapes has not been explored. Detailed analysis about the propagation of coherences (off-diagonal elements) in both slow and fast modulation limit has also not been performed.

Experimentally first Zewail *et. al.*[23-25] and subsequently Fayer and co-workers[26,27] observed presence of quantum coherence in early eighties. Haken *et. al.*[28-30] and independently Silbey and co-workers[31,32] calculated absorption line shape in the presence of exciton transfer in dimer. Their study was limited to the case where he line shape was Lorentzian. Later Sumi[33] explained Gaussian and Lorentzian behavior of the line shape in strong scattering and weak scattering limit. Silbey *et. al.*[34] further investigated the exciton transfer line shape which extended Sumi's observation. The main limitation of these studies was the restriction of the assumption that bath fluctuations obeyed Markov process. In an interesting study, Scholes *et. al.*[35,36] experimentally detected splitting of absorption band of dimer naphthalene, although the absorption bands were not symmetric due to the contribution of vibration. Jang and Silbey[37] assumed quasi-static disorder and calculated line shape for model dimer and B850 ring of LH2 photosynthetic complex using quantum master equation based formalism. Schröder *et. al.*[38] considered static disorder and compared line shape of B850 ring of LH2. Most of the above studies employed either Markovian approximation or used an effective non-Markovian theory where dynamic disorder was not considered explicitly.

A few years ago Donehue *et. al.*[39] studied cyclic thiophene oligomers of 12, 18, 24 and 30 repeating unit. From three pulse photon echo shift measurement they concluded that with increasing ring size, coupling to the bath weakened whereas intra-molecular coupling increased.

As the present study focuses on optical line shape, we first briefly discuss the pioneering work of Kubo[40,41]. This work of Kubo created the language we use in the discussion of line shapes. The analysis is general and was originally developed to explain NMR line shape.[42-44]

Kubo's analysis starts with a simple stochastic equation of motion for the coordinate x(t) of a harmonic oscillator coupled with its environment,

$$\dot{x}(t) = i\omega(t)x(t) \tag{1}$$

where x(t) is the harmonic oscillator co-ordinate and $\omega(t)$ is the frequency modulation by the interaction with the environment. Time dependent frequency is written as

$$\omega(t) = \omega_0 + \delta\omega(t) \tag{2}$$

In Kubo formalism, the spectral line shape is defined as

$$I(\omega) = \frac{1}{2\pi}\int_{-\infty}^{\infty} e^{-i\omega t}\left\langle \exp\left(i\int_0^t \delta\omega(s)\,ds\right)\right\rangle dt \tag{3}$$

Where < … > implies an averaging over a time trajectory, that is a time average. When the frequency modulation is a Gaussian Markov stationary process, the cumulant beyond second order vanishes. If the time correlation function of the frequency modulation is exponential in time, the integral in the exponent can be carried out and the final expression is well-known and is given by

$$\phi(t) = \exp\left[ -\Delta^2 \tau_c^2 \left\{ \exp\left( -\frac{t}{\tau_c} \right) + \frac{t}{\tau_c} - 1 \right\} \right] \tag{4}$$

where $\Delta$ is the strength of the random modulation and $\tau_c$ is the relaxation time.

For slow modulation case when $\tau_c$ is large, line shape is Gaussian and line shape is termed inhomogenously broadened. For fast modulation case time correlation function is exponential and line shape is Lorentzian which is often, as mentioned above, is referred to as the motional narrowing limit. As the cumulant expansion beyond 2$^{nd}$ order vanishes one may think that the above approach is correct but this approach may fail even for Gaussian bath, for the following reason. In the interaction representation (Eq. (7)), the full interaction is quantum mechanical and proper time ordering is required to truncate the cumulant expansion. Thus, even when the classical bath is Gaussian, the cumulant expansion may not be truncated trivially at second order. The main point here is that the quantum stochastic Liouville equation does not suffer from such limitations. And it can be implemented exactly for two state Poisson bath. The implementation of QSLE for Gaussian bath can in principle can also be done exactly (as shown in Refs. 19,45 ). However its implementation is difficult. Because one needs to include at least 6 or 7 excited bath states which makes even a numerical approach prohibitively difficult. Nevertheless, QSLE provides an exact approach to the problem of quantum-classical mixed problems we encounter in spectroscopy.

It is straightforward to apply the QSLE-based Kubo lineshape formalism even when we have multiple energy states participating in the dynamical process, such as in a dimer. However, the solution is now non-trivial. The reason is the increase in the number of fluctuation sources. For a dimer, we can now have two diagonal fluctuations and one off-diagonal fluctuation terms. In addition, there is usually a constant off-diagonal term in the dimer. The constant off-diagonal term give rise to pronounced oscillations produced by quantum coherence between the two states.

As discussed by several authors, there are two well-known physical systems where coherence in population transfer dynamics could play important role. These are (i) thin films of conjugated polymers, and (ii) photosynthetic reaction center. In both these systems, interplay between off-diagonal coupling *J* and fluctuation or dynamic disorder can limit the extent of coherence. Note that even in the absence of time dependent fluctuation, one can have static disorder in off-diagonal coupling. This could be particularly prevalent in thin films of conjugated polymer.

The initial theories of exciton transfer of Haken-Strobl-Reineker and of Silbey *et. al.* are based on the following standard (often referred to as Haken-Strobl) Hamiltonian

$$H_{tot} = H_S + H_B + H_{\text{int}} \tag{5}$$

where system (exciton) Hamiltonian is defined as

$$H_S = \sum_k \omega_0 |k\rangle\langle k| + \sum_{\substack{k,l \\ k\neq l}} J_{kl} |k\rangle\langle l| \quad . \tag{6}$$

where $\omega_0$ is the energy of an electronic exciton localized at site $k$ and $J_{kl}$ is the time uncorrelatedoff-diagonal interaction between excitations at site $k$ and $l$ . $H_B$ is the bath Hamiltonian and $H_{\text{int}}$ is the interaction Hamiltonian between the system and the bath. The Hamiltonian can be represented as follows,

$$H(t) = H_S + V(t) \tag{7}$$

where, setting $\hbar = 1$, $V(t)$ has the following form

$$V(t) = e^{iH_B t} V_{class} e^{-iH_B t} . \tag{8}$$

Thus in the interaction representation, the coupling potential is time dependent which we write as

$$V(t) = \sum_{k} |k\rangle\langle k| V_d(t) + \sum_{\substack{k,l \\ k \neq l}} |k\rangle\langle l| V_{od}(t) \quad . \qquad (9)$$

Here $V_d(t)$ denote diagonal (local) and whereas $V_{od}(t)$ represents off-diagonal (non-local) parts of the fluctuating potential $V_{int}$. In this work we model $V$ (t) by a stochastic function with known statistical properties. Using the above Hamiltonian Bagchi and Oxtoby,[45] as well as Dutta and Bagchi[46] studied exciton diffusion in a one dimensional system of regularly places two level systems, both in the continuum and in the discrete limits, respectively. We assume both the fluctuation to be described by Poisson stochastic process. In fact, the Poisson bath can be considered as a limiting form of Gaussian bath. For Poisson bath case, diagonal and off-diagonal matrix elements jump between two values ( $\pm V$ ) such that the average of each matrix element is zero. For most of the calculations reported here, we have, for simplicity, neglected the off-diagonal fluctuation. For real systems (photo-synthetic systems) off-diagonal fluctuation is often small with compared to the diagonal one. However, we have considered an interesting role of off-diagonal fluctuation in exciton transport mechanism.

In this work we are primarily interested in study of quantum coherence that could be created through optical line shapes in extended systems. In particular, we explore the correlation between optical line shape and quantum coherence in the extended system composed of equally spaced two levels. We also consider cyclic polymers composed of the correlated two level systems.

The main objectives of the present works are the followings: (1) to study of the line shape for excitation transfer in discrete model systems in slow and fast modulation limit, (2) to study the effects of diagonal fluctuation for both correlated and uncorrelated bath case on the line shape differ for linear and cyclic systems, (3) to connect population transfer dynamics with line shape in both slow modulation and fast modulation limit for both linear and cyclic systems, (4) to perfume a detailed analysis of the propagation of coherence for correlated and uncorrelated bath case in slow and fast modulation limit.

We obtain nearly the same results when all the baths are fully correlated (a limit we refer to as the correlated bath) as when they are completely uncorrelated, in fast modulation limit. In this limit both line shape and OPF behave in similar manner. However, in the slow modulation limit, the spatial correlation is important. In the latter case the line shapes as well as population transfer dynamics for the correlated bath and uncorrelated bath are significantly different. We attribute this to the effects of quantum coherence affecting the two limits differently.

The organization of the rest of the paper is as follows: In Sec. II we explain Kubo's quantum stochastic Liouville equation (QSLE). In Sec. III we discuss eigen value and eigen vector of Toeplitz matrix to explain line shape behavior in fast modulation limit. In Sec. IV we have described calculation of line shape function for monomer. In Sec. V we elucidate line shape calculation for multichromophoric systems. In Sec. VI, we illustrate occupation probability function using QSLE. In Sec. VII we explore the propagation of coherence for correlated and uncorrelated bath case (diagonal fluctuation). In Sec. VIII we discuss the limitation of QSLE and effect of temperature on line shape and exciton transfer dynamics. Finally, in Sec. IX we draw the conclusion. Coupled equation of motion for dimer system in case of correlated bath can be found in **APPENDIX.**

## II. QUANTUM STOCHASTIC LIOUVILLE EQUATION

Quantum stochastic Liouville equation (QSLE) was derived by Kubo to incorporate stochastic fluctuations in quantum Liouville equation. QSLE is ideally suited to describe dynamics of a quantum system interacting with a classical bath. It is extensively applied to study electron spin resonance (ESR) and nuclear magnetic resonance (NMR) studies as well as vibrational relaxation (for both energy and phase). It was earlier to successfully address many aspects of exciton migration in a system of two level systems in the presence of dynamic disorder.

There are several studies of such coupled quantum system-classical bath systems.[47-50] Skinner and co-workers[51,52] made a notable contribution in this area in the context spectroscopy. Hernandez and Voth[53] explored the role of classical mechanics in defining coherence in quantum system through the calculation of time correlation function. Bagchi and Oxtoby studied exciton transport in one dimensional lattice considering correlated bath and uncorrelated bath for continuum model using QSLE. It was shown in the latter work that population transfer dynamics is sensitive to the nature of bath fluctuations in the non-Markovian limit.

We start with the quantum Liouville equation,

$$\frac{\partial \rho}{\partial t} = -\frac{i}{\hbar}\left[H(t), \rho\right]. \tag{10}$$

where the Hamiltonian is given by Eq. (7). The master equation for probability measure is provided as follows,

$$\frac{\partial W(\mathrm{V}, t)}{\partial t} = \Gamma_{\mathrm{V}} W(\mathrm{V}, t). \tag{11}$$

where $\mathrm{V}$ is random variable, $W(\mathrm{V}, t)$ is the probability measure and $\Gamma_{\mathrm{V}}$ is the stochastic diffusion operator. Now a joint probability distribution $P(\rho, V, t)$ is defined as follows

$$P(\rho, \mathrm{V}, t) = \left\langle \delta\left(\rho - \rho(t)\right) \delta\left(\mathrm{V} - \mathrm{V}(t)\right) \right\rangle. \tag{12}$$

In the next step, one derives an equation of motion for the joint probability distribution. We shall forego the details as they have described in many places. For the present work, we define a reduced density matrix $\sigma(t)$ as,

$$\sigma(t) = \int d\rho \, \rho \, P(\rho, \mathrm{V}, t) . \qquad (13)$$

Kubo showed that the reduced density matrix follow the following modified, stochastic quantum Liouville equation

$$\frac{\partial \sigma}{\partial t} = -\frac{i}{\hbar}\left[H(t), \sigma\right] + \Gamma_{\mathrm{v}} \sigma . \qquad (14)$$

Eq. (14) is known as Kubo's stochastic Liouville equation. The above equation can be used to study the energy transfer dynamics of the system. A different form of the equation is used to obtain the line shape function.[44] In this study we have considered both correlated and uncorrelated bath case. For correlated bath case all the diagonal l and off-diagonal fluctuation are spatially correlated all the time. On the other hand, for uncorrelated bath case all the diagonal and off-diagonal fluctuations are completely uncorrelated.

## III. USEFUL PROPERTIES OF TRIDIAGONAL TOEPLITZ MATRIX : EIGENVALUES AND EIGENVECTORS TO ANALYZE NATURE OF THE LINE SHAPE IN FAST MODULATION LIMIT

The study of coherence in our extended system of an array of two levels becomes simplified considerably by the use of certain properties from matrix algebra. In the absence of any fluctuation in the off-diagonal and diagonal terms, our system Hamiltonian with constant off-diagonal coupling J is a tridiagonal Toeplitz matrix[54-56] in the site representation for linear system. This leads to several non-trivial observations. First, we have an analytic closed form expression for the eigenvalues and eigenvectors of the system. If we denote the diagonal terms by *J*, then the tridigonal matrix is given by

$$\begin{pmatrix} \omega_0 \; J \; 0\;0\;0\;0\;0 \ldots\ldots\ldots \\ J \; \omega_0 \; J \; 0\;0\;0\;\;0 \ldots\ldots\ldots \\ 0 \; J \; \omega_0 \; 0\;0\;0 \ldots\ldots\ldots \\ 0\;0\; J \; \omega_0 \; J \; 0\;0\;0 \ldots\ldots \\ \ldots\ldots\ldots\ldots\ldots\ldots\ldots\ldots\ldots \\ \ldots\ldots\ldots\ldots\ldots\ldots\ldots\ldots J \; \omega_0 \end{pmatrix} \tag{15}$$

Another remarkable property of this tridiagonal Toeplitz matrix is that all the tridiagonal Toeplitz matrices have same eigenvectors.

Another important result is that for $N \to \infty$, i.e. for infinite number of sites for linear and cyclic models the eigen values are bounded. For both linear and cyclic system maximum and minimum eigen values are $\omega_0 \pm 2J$. Consequently for infinite size system we obtain dense spectrum with large number of peaks.

In the fast modulation limit when the rate of fluctuation is high and amplitudes of diagonal and off-diagonal coupling constants are small, the unperturbed system Hamiltonian can be used to analyze the nature of the line shape, particularly the positions of the peaks of the spectrum. In this case system Hamiltonian is a Toeplitz matrix. We note in passing that for cyclic models the matrices are known as circulant. The eigen value of Toeplitz matrix for linear and cyclic models give positions of the line shape peaks. Square of the co-efficients in each eigen states is related to the intensity of each peak. For linear and cyclic models the expression of eigen value and co-efficient which is required to obtain the eigen states are provided as follows,

### A. Linear models

Here the eigen values are given by [**54-56** ]

$$E_j = \omega_0 + 2J\cos\frac{j\pi}{N+1} \qquad \text{(eigen values)}$$

(16)

$$C_{jk} = \left(\frac{2}{N+1}\right)^{1/2} \sin\frac{jk\pi}{N+1} \qquad \text{(coefficients for eigen vectors)}$$

(17)

where $N$ is total number of site, $j = 1, 2, 3, ......, N$ and also $k = 1, 2, 3, ......, N$ . Note the following important points. First, the eigenvalues are bounded between $\omega_0 + 2J$ and $\omega_0 - 2J$ . Therefore, in the large N limit, the eigen values form almost a continnum. This is an important result as it suggests that the spectrum can be broad in the large N limit but bounded between the two limits given above. Second, the eigen functions can be easily found in an analytical form, from equation (17). Thus, for a trimer where the unperturbed basis set are localized site functions, the expression for the eigen states is as follows

$$\begin{aligned}
\psi_1 &= 0.5\phi_1 + 0.71\phi_2 + 0.5\phi_3, \quad with\, E_1 = \omega_0 + J/\sqrt{2} \\
\psi_2 &= 0.71\phi_1 - 0.71\phi_3, \quad with\, E_2 = \omega_0 \\
\psi_3 &= 0.5\phi_1 - 0.71\phi_2 + 0.5\phi_3, \quad with\, E_3 = \omega_0 - J/\sqrt{2}
\end{aligned}$$

(18)

where $\psi_i$ denotes the $i^{th}$ eigen state of the system with eigenvalue $E_j$, and $\phi_i$ denotes $i^{th}$ molecular states without any intersite coupling.

The eigenvalues give the positions of the absorption maxima in the motional narrowing limit. There can be additional maxima in the case of static modulation, as we show later. Co-efficient of eigen vectors are related to the intensity of the peaks. For all the linear systems with odd number of sites the centered peak has large intensity than the others and. For the model consists of an even number of sites, the two peaks in the center have same intensity but greater than all other peaks. For cyclic system, the situation is different and rather interesting. Here the intensity of the low frequency peaks than that those at high frequency, due to the degeneracy of states.

Since the eigen vectors of our Toeplitz matrix are fully known, we can easily write down expressions for the wave functions of larger systems.

**B. Cyclic models**

In this case also we have analytical expressions of eigenvalues and eigen vectors that are given by similar expressions

$$E_j = \omega_0 + 2J\cos\frac{2\pi(j-1)}{N} \quad \text{(eigen values )} \tag{19}$$

$$C_{jk} = \frac{1}{\sqrt{N}}\exp\left[\frac{2\pi i(k-1)(j-1)}{N}\right] \quad \text{(coefficients for eigen vectors)} \tag{20}$$

where $N$ is total number of site, $j = 1, 2, 3, ......, N$ and also $k = 1, 2, 3, ......, N$ .

For infinite number of site eigen value attain the maximum.

The peak positions obtained from QSLE equation are exactly the same as the ones given above.

For cyclic trimer, the lowest eigenvalue corresponds to doubly degenerate state. All the eigen states are provided as follows,

$$\begin{aligned}
\psi_1 &= 0.58\left(\phi_1 + \phi_2 + \phi_3\right) \\
\psi_2 &= 0.82\phi_1 - 0.41\phi_2 - 0.41\phi_3 \\
\psi_3 &= 0.82\phi_1 + 0.71\phi_2 - 0.71\phi_3
\end{aligned} \tag{21}$$

In this case, intensity of the first peak is greater than that of last peak. Due to degeneracy we obtain only two peaks.

# IV. CALCULATION OF LINE SHAPE FUNCTION

## A. MONOMER

We present a study of monomer with QSLE as this is not only the simplest system (as only the site energy is subjected to the fluctuating environment) but also it helps bringing certain important aspects, as discussed below. We note that this is the same system as the one for studied by Kubo long time ago to initiate systematic study of line shape problems, as discussed in Sec. I. Kubo's treatment however was phenomenological while the treatment given below follows a more accurate alternative derivation (again by Kubo and discussed above). Here we start with the following definition of the line shape function[44]:

$$I(\omega) = \frac{\mu^2}{\pi} \mathrm{Re}\left\langle 0 \middle| X(i\omega) \middle| 0 \right\rangle \tag{22}$$

where $\dot{X}(t) = iHX + \Gamma X$ . (23)

Here the whole quantity inside the bra and ket is similar as Fourier transform of transition dipole moment dipole moment correlation function.

In the case of the Poisson bath, one can write the stochastic bath operator $\Gamma$ and the eigen vectors as follows,

$$\Gamma = \begin{pmatrix} -\frac{b}{2} & \frac{b}{2} \\ \frac{b}{2} & -\frac{b}{2} \end{pmatrix}; \quad |0\rangle = \begin{pmatrix} 1 \\ 1 \end{pmatrix}; \quad \langle 0| = \begin{pmatrix} \frac{1}{2} & \frac{1}{2} \end{pmatrix} \tag{24}$$

We consider the ground state energy as zero. Then the Hamiltonian consists of energy of the excited state and diagonal fluctuation due to interaction of system with bath. Line shape function is given as

$$I(\omega) = \frac{\mu^2}{\pi} \frac{b_d \left[ V_d^2 + (\omega - \omega_0)^2 \right]}{\left[ V_d^2 + (\omega - \omega_0)^2 \right]^2 + 4b_d^2 (\omega - \omega_0)^2} \tag{25}$$

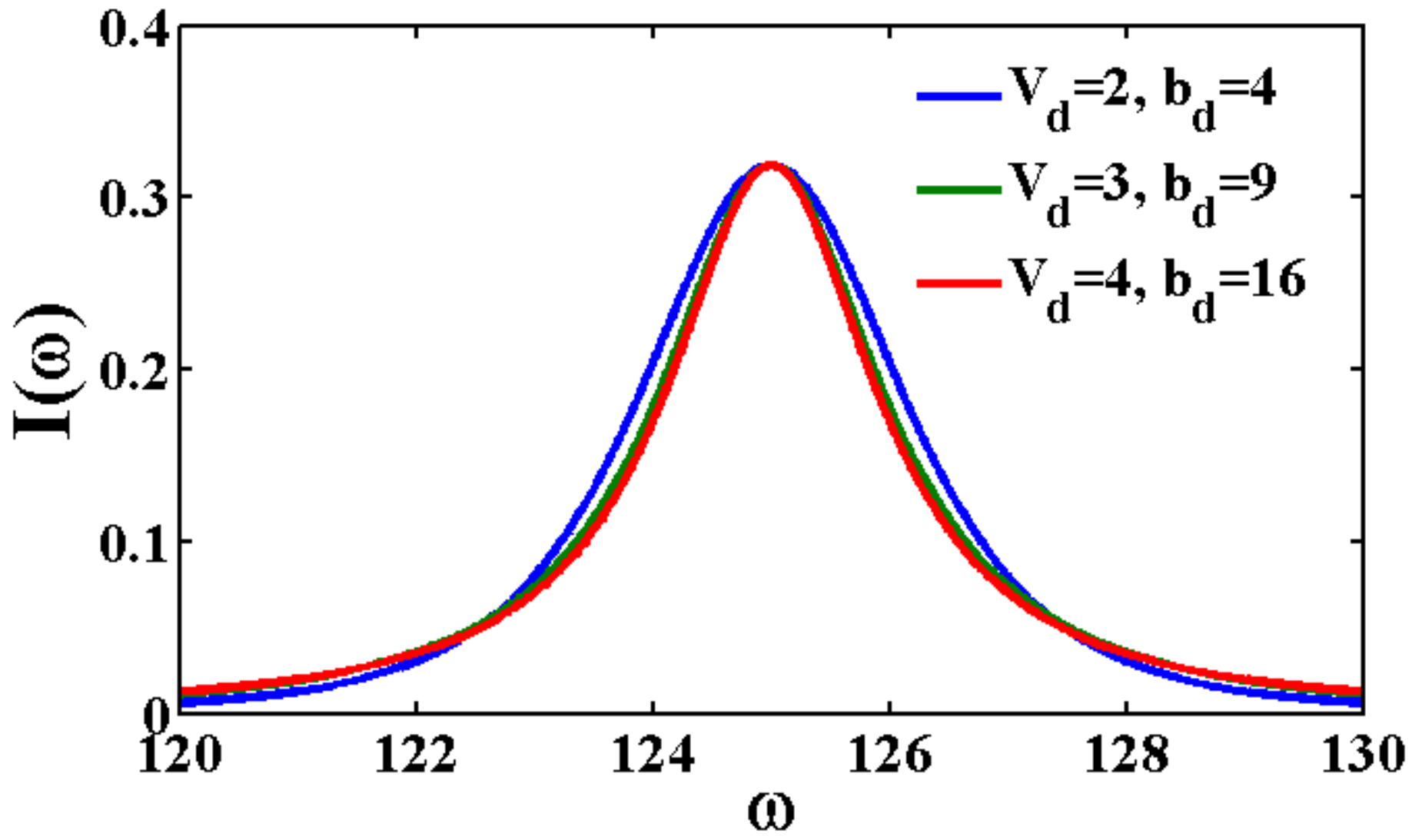

**Fig. 1. Line shape function is plotted against frequency for different value of diagonal rate of fluctuation. For high value of $b_d$ the line shape shows sharp Lorentzian behavior and for low value of $b_d$ the line shape broadened and intensity decreases.**

Fig. 1 shows line shape for monomer. In this case $\omega$ is scaled by $V_d^2 / b_d$. We have kept the ratio $V_d^2 / b_d$ fixed. Note that the change in the nature of line shape is quite small as maximum intensity of the peak I s $\sim \frac{b_d}{V_d^2}$ . For high values of $V_d$ and $b_d$ , the line shape approaches Kubo predicted line shape in the motional narrowing limit. That is, the line shape is sharply Lorentzian at peak position. However, with decreasing $b_d$ , although the line shape gets broadened with low intensity, it does not approach the Gaussian behavior which is different from the prediction of Kubo's phenomenological theory that predicts a Gauusian line shape in the slow modulation limit. This is probably due to the following reason. In Kubo's case the bath was Gaussian and for that case the cumulant expansion beyond 2$^{nd}$ order is zero though the method cannot be fully justified (Sec I). In the case of Poisson bath all the moments exist and truncation after second order does not provide exact results. Note however that in the QSLE approach that we have used does not require such approximation, and the line shape we obtain is exact for Poisson bath for the chosen Hamiltonian.

## B. Line shape of multichromophoric system

Definition of Kubo's[23] line shape function can be written as

$$I(\omega)=\frac{1}{\pi}\mathrm{Re}\sum_{k,l}\mu_k\left\langle 0\right|X_{kl}(i\omega)\left|0\right\rangle\mu_l \tag{26}$$

If the transition dipole moments couples ground state with only one excited state (say 1$^{st}$ excited state or excited state for site 1), we can write $\mu_1 \neq 0$ whereas all other elements are zero.

Subsequently Eq. (26) is reduced to the following expression

$$I(\omega)=\frac{\mu_1^2}{\pi}\mathrm{Re}\left\langle 0\right|X_{11}(i\omega)\left|0\right\rangle \quad , \tag{27}$$

where $\dot{X}(t)=iHX+\Gamma X$ . (28)

We have only considered 1st element of the matrix to get the line shape function as transition dipole moment only couples ground states to one of the excited state and also considered later the transition diploe moment as unity.

Laplace transform of Eq. (28) gives

$$(s - iH - \Gamma)X[s] = \mathbf{1} \tag{29}$$

### (i) Dimer system

For a dimer, the Hamiltonian can be written generally as,

$$H = \begin{pmatrix} \omega_0 + V_d(t) & J + V_{od}(t) \\ J + V_{od}(t) & \omega_0 + V_d(t) \end{pmatrix} \tag{30}$$

The diagonal and off-diagonal fluctuations can take only two values $\pm V_d$ and $\pm V_{od}$. In the subsequent discussion in this section, we shall set off-diagonal fluctuation as zero. For real systems like photosynthetic system the value of off-diagonal fluctuations is quite low compare to the diagonal fluctuation and other parameters present in the Hamiltonian.

For dimer model if we consider transition dipole moment only couples 1st excited state to ground state. Now, one can write Eq. (29) as follows

$$\begin{aligned} \left(s - i\left(\omega_0 + \mathbf{V_d}\right) - \mathbf{\Gamma}\right)\mathbf{x_1} - iJ\mathbf{x_2} = \mathbf{x_1^0} \\ \left(s - i\left(\omega_0 + \mathbf{V_d}\right) - \mathbf{\Gamma}\right)\mathbf{x_1} - iJ\mathbf{x_2} = \mathbf{x_2^0} \end{aligned} \quad , \tag{31}$$

where $\mathbf{x_1}$ and $\mathbf{x_2}$ are the components of X i.e. first and second excited state. For dimer system 3 states are possible when both the chromophores are in ground state and one is in excited whereas other one is in ground states (two states are possible). We assume that transition dipole moment only couples one of the excited states to the ground state. Solving Eq. (31) for $\mathbf{x_1}$ with the condition $\mathbf{x_1^0} = \mathbf{1}$ and $\mathbf{x_2^0} = \mathbf{0}$, we obtain the element $X_{11}[i\omega]$ in the form

$$X_{11}[i\omega] = (\mathbf{AA} - \mathbf{BB})^{-1}\mathbf{A} \tag{32}$$

where $\mathbf{A} = \begin{pmatrix} i\omega - i\omega_0 - iV_d + \frac{b}{2} & -\frac{b}{2} \\ -\frac{b}{2} & i\omega - i\omega_0 + iV_d + \frac{b}{2} \end{pmatrix}$ and $\mathbf{B} = \begin{pmatrix} -iJ & 0 \\ 0 & -iJ \end{pmatrix}$ (33)

We now discuss the two limiting cases for bath fluctuation, one fully correlated and the other fully uncorrelated.

**(a) Correlated bath**

Hence the expression for line shape function is given by

$$I = \frac{1}{2\pi}\mathrm{Re}\left[\frac{PQ^2 + PJ^2 + 2R^3 - 2PQR + 2J^2R - PR^2 - QR^2 + P^2Q + J^2Q}{(P^2 + J^2 + R^2)(Q^2 + J^2 + R^2) - (PR + RQ)^2}\right] \tag{34}$$

where $P = \left(i\omega - i\omega_0 - iV_d + \frac{b}{2}\right)$, $Q = \left(i\omega - i\omega_0 + iV_d + \frac{b}{2}\right)$, $R = -\frac{b_d}{2}$ (35)

For photo synthetic systems, spectroscopic measurements give $\omega_0 \sim 12500\,\mathrm{cm}^{-1}$ and $J \sim 100 - 300$ cm$^{-1}$. We consider $\omega_0 = 125$ where it is scaled by $J$ and $J = 1$. The variable $\omega$ is also scaled by $J$.

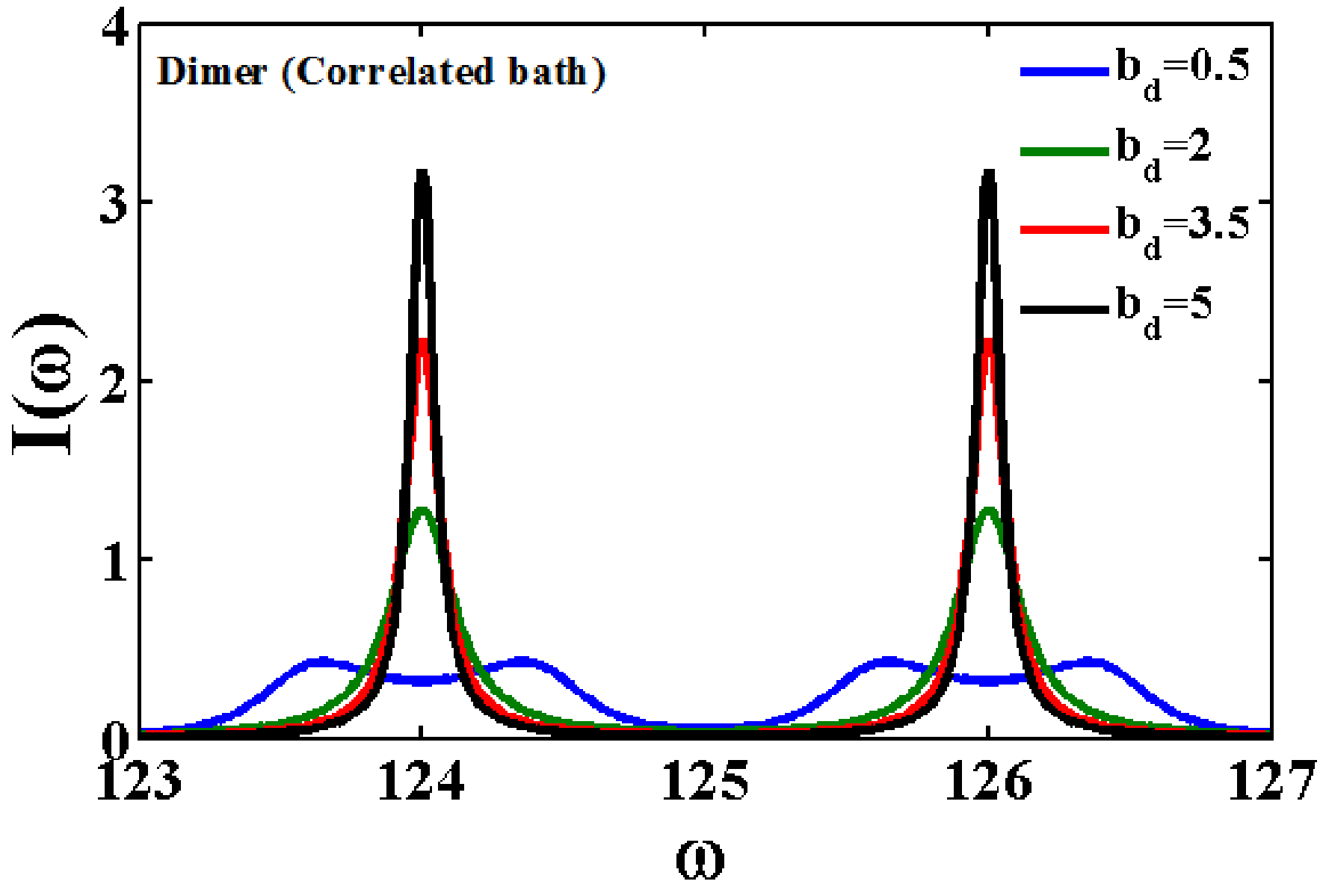

**Fig. 2. Line shape function is plotted against frequency for dimer system J=1, $V_d$=0.5 for correlated bath. Only $b_d$ varies from low value to high value. Going from slow modulation to fast modulation limit multiple peaks are merged with each other to show sharp narrowed peaks.**

### (b) Uncorrelated bath

Calculation of line shape of uncorrelated bath case is quite complicated as one has to consider all the contribution from each of the uncorrelated bath. We only provide plots of the line shape function.

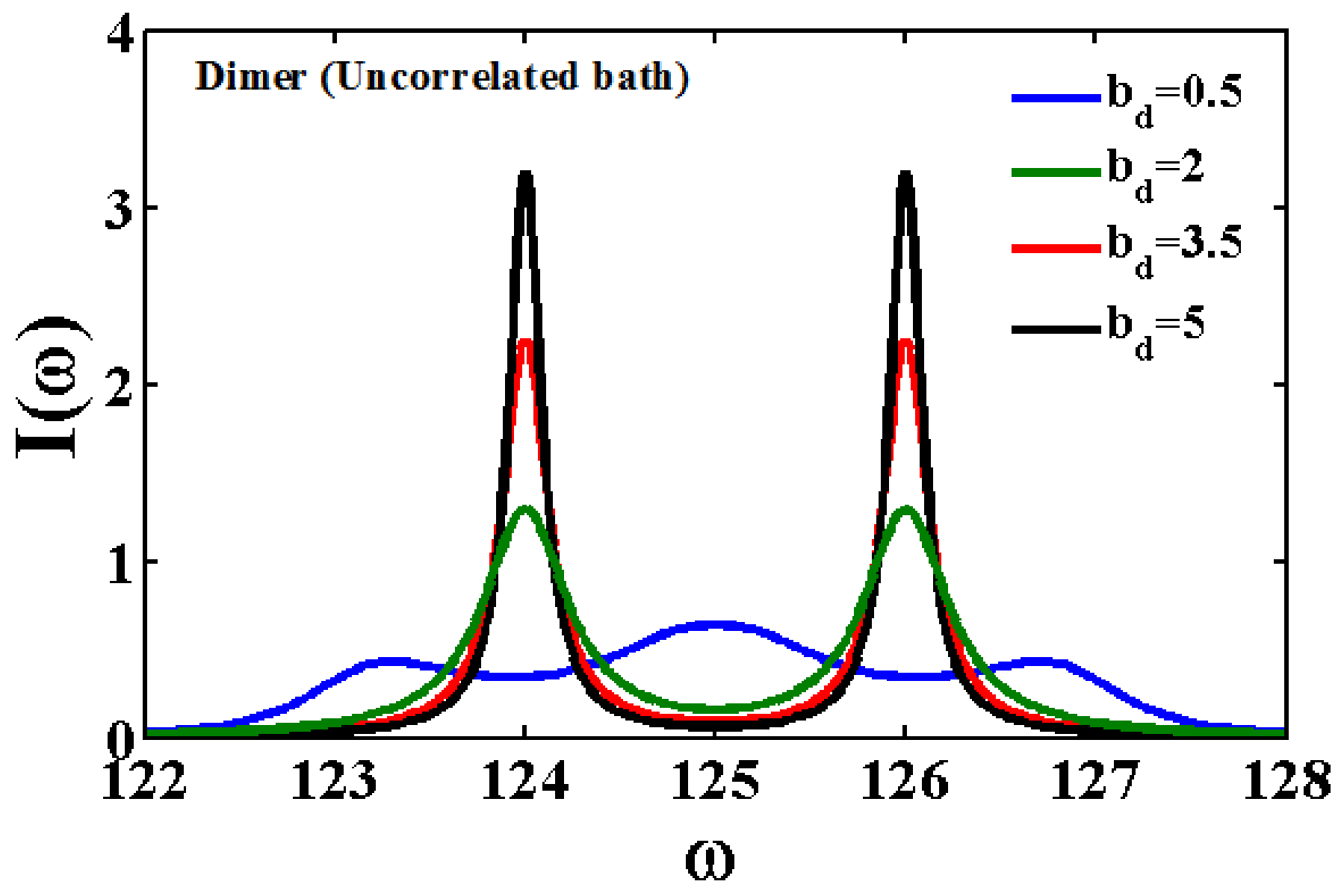

**Fig. 3. Line shape function is plotted against frequency for dimer system J=1, $V_d$=0.5 for uncorrelated bath. Only $b_d$ varies from low value to high value. In slow modulation limit number of peaks are less than that of correlated bath case and peaks are broadened. In case of fast modulation limit intensity and broadening of all the peaks are greater than that of correlated bath case.**

### (ii) Linear tetramer

Similarly for linear 4 site model we can write Eq. (29) as follows

$$\begin{aligned} &\mathbf{Ax}_1 + \mathbf{Bx}_2 = \mathbf{1} \\ &\mathbf{Bx}_1 + \mathbf{Ax}_2 + \mathbf{Bx}_3 = \mathbf{0} \\ &\mathbf{Bx}_2 + \mathbf{Ax}_3 + \mathbf{Bx}_4 = \mathbf{0} \\ &\mathbf{Bx}_3 + \mathbf{Ax}_4 = \mathbf{0} \end{aligned} \tag{36}$$

By solving the above Eq. (36) we get the expression of $X_{11}[i\omega]$ as follows

$$X_{11}[i\omega] = \frac{1}{2}\left[(\mathbf{A}+\mathbf{B})\mathbf{A}-\mathbf{BB}\right]^{-1}(\mathbf{A}+\mathbf{B}) + \frac{1}{2}\left[(\mathbf{A}-\mathbf{B})\mathbf{A}-\mathbf{BB}\right]^{-1}(\mathbf{A}-\mathbf{B}) \tag{37}$$

**A** and **B** are already given for the case of two site model.

#### (a) Correlated bath

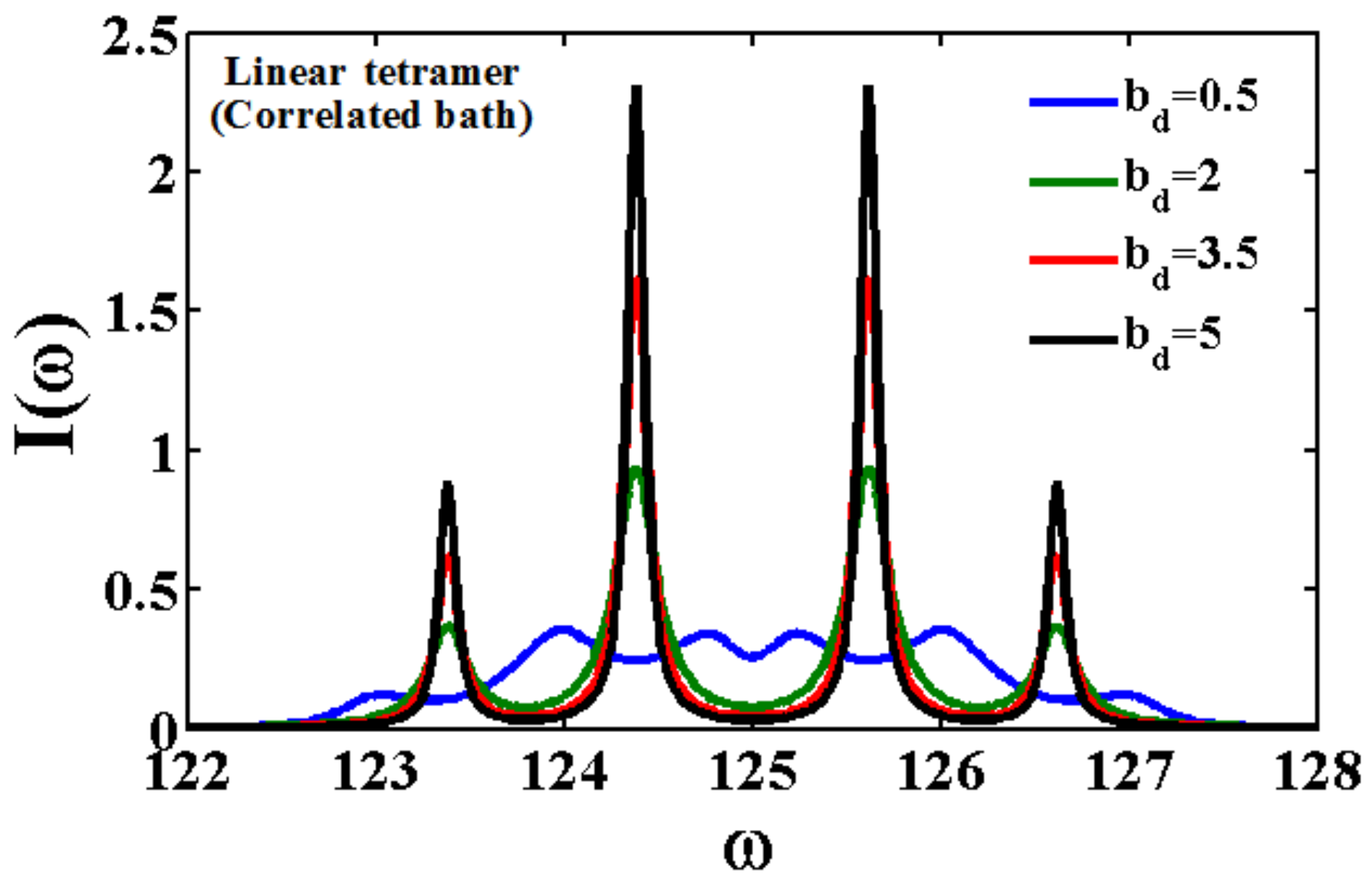

**Fig. 4. Line shape function is plotted against frequency for linear tetramer system J=1, $V_d$=0.5 for correlated bath. Only $b_d$ varies from low value to high value. With increasing the rate of fluctuation, number of peaks decreases and all the peaks suggests motional narrowing behavior. Maximum intensity of the peaks are larger for dimer system than that of linear tetramer.**

**(b) Uncorrelated bath**

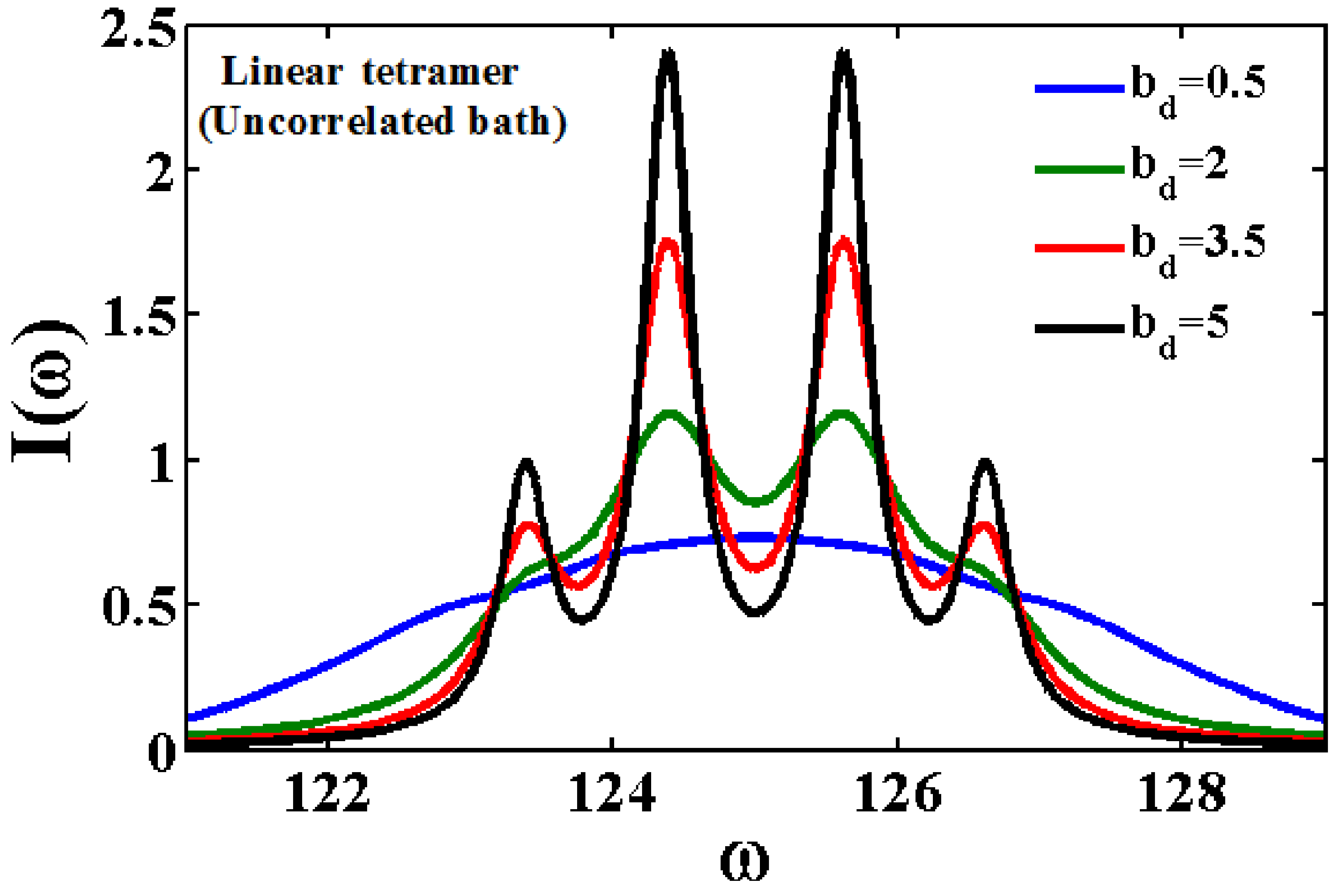

**Fig. 5. Line shape function is plotted against frequency for linear tetramer system J=1, $V_d$=0.5 for uncorrelated bath. Only $b_d$ varies from low value to high value. In slow modulation limit peaks**

**broadened and with increasing the modulation rate splitting of the peaks are taking place. Motional narrowing limit can be obtained by increasing $b_d$ to a very large value (greater than that of uncorrelated bath case.**

### (iii) Cyclic tetramer

Similarly for cyclic 4 site model we can write Eq. (29) as follows

$$\begin{aligned} \mathbf{Ax_1} + \mathbf{Bx_2} + \mathbf{Bx_4} &= \mathbf{1} \\ \mathbf{Bx_1} + \mathbf{Ax_2} + \mathbf{Bx_3} &= \mathbf{0} \\ \mathbf{Bx_2} + \mathbf{Ax_3} + \mathbf{Bx_4} &= \mathbf{0} \\ \mathbf{Bx_1} + \mathbf{Bx_3} + \mathbf{Ax_4} &= \mathbf{0} \end{aligned} \tag{38}$$

By solving the above Eq. (38) we get the expression of $X_{11}[i\omega]$ as follows

$$X_{11}[i\omega] = \frac{1}{2}\left[(\mathbf{A}+\mathbf{B})(\mathbf{A}+\mathbf{B}) - \mathbf{BB}\right]^{-1}(\mathbf{A}+\mathbf{B}) + \frac{1}{2}\left[(\mathbf{A}-\mathbf{B})(\mathbf{A}-\mathbf{B}) - \mathbf{BB}\right]^{-1}(\mathbf{A}-\mathbf{B}) \tag{39}$$

#### (a) Correlated bath

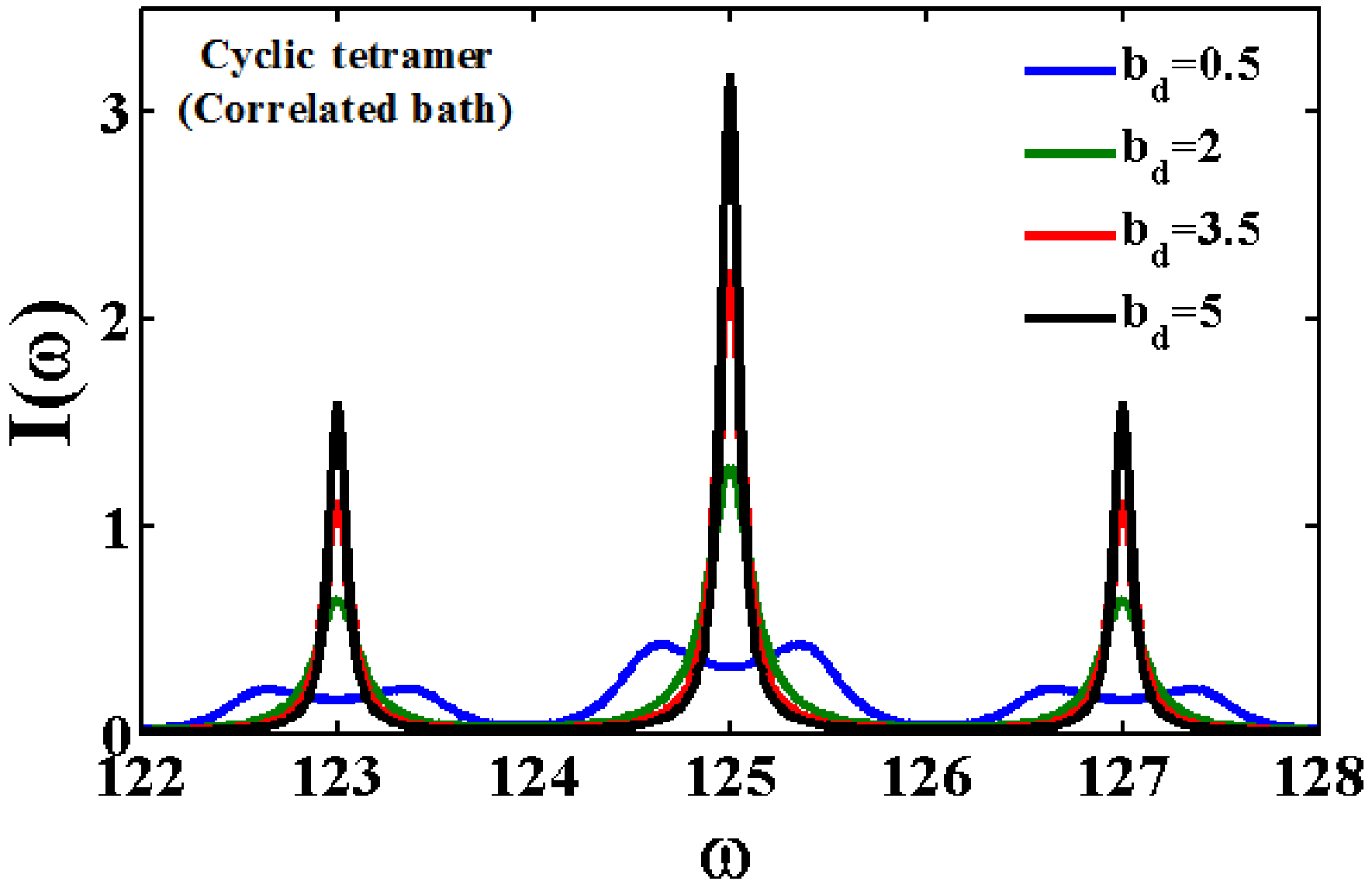

**Fig. 6. Line shape function is plotted against frequency for cyclic tetramer system J=1, $V_d$=0.5 for correlated bath. Only $b_d$ varies from low value to high value. Going from slow modulation to fast modulation limit each pair of peaks merge with each other to produce a single peaks. Maximum intensity of peaks for cyclic system is greater than that linear system with same number of sites.**

**(b) Uncorrelated bath**

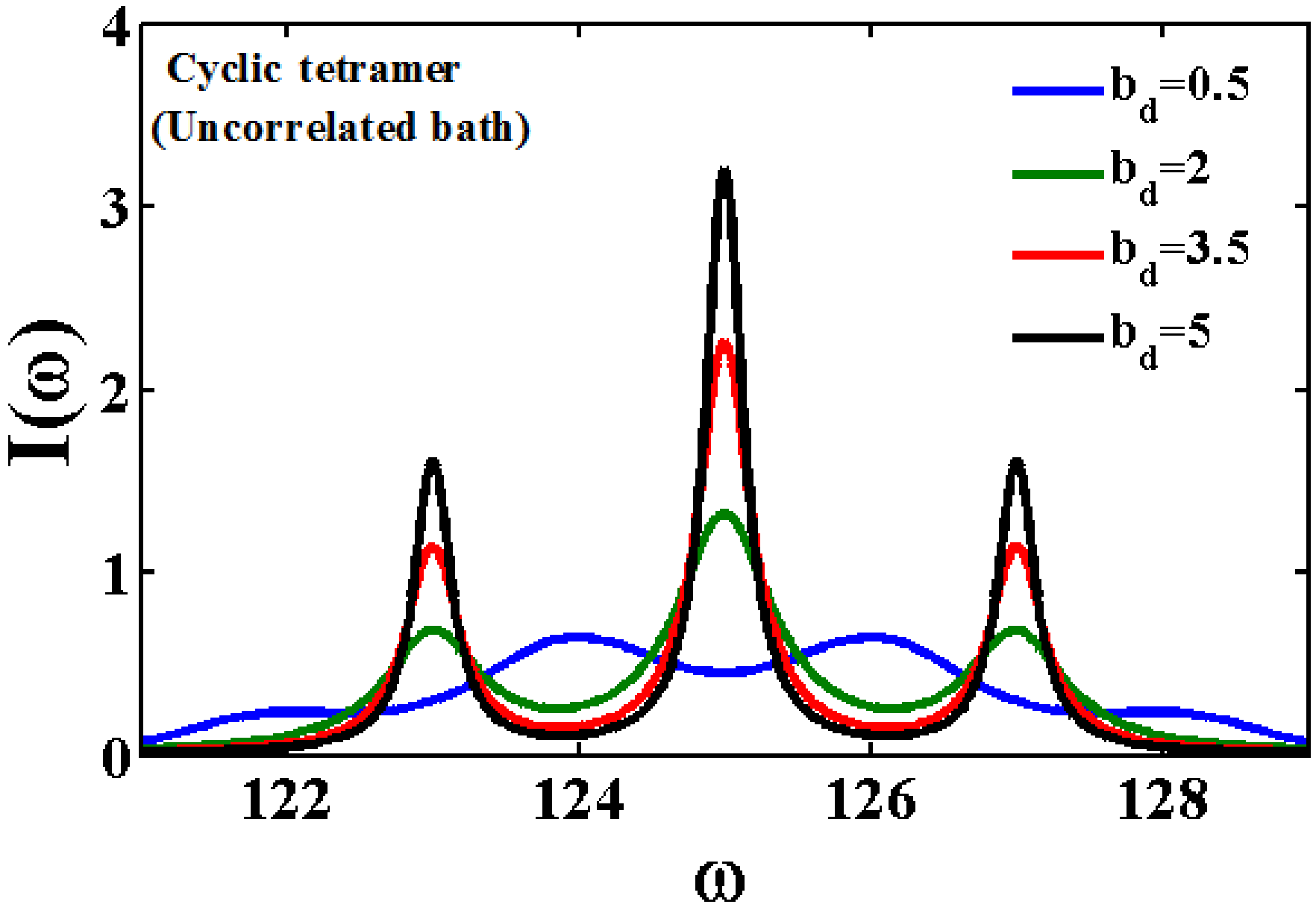

**Fig. 7. Line shape function is plotted against frequency for cyclic tetramer system J=1, $V_d$=0.5 for uncorrelated bath. Only *$b_d$* varies from low value to high value. In slow modulation limit peaks are broadened and with increasing the rate of fluctuation, intensity of peaks increase and shows the narrowing behavior.**

In **Figures 2-7** we have plotted line shapes for both correlated and uncorrelated bath for dimer, linear tetramer and cyclic tetrameric systems. *For both correlated and uncorrelated bath, the intersite coupling J act as an off-diagonal perturbation and is responsible for the splitting of the peaks.* For linear system the number of peaks is equal to the number of sites, as given by the Toeplitz matrix. For cyclic systems, number of peaks is accordance with the cyclic symmetry (that is, number of sites minus unity for cyclic tetramer). Interestingly, however, further splitting of the peak occurs with lowering the rate of fluctuation $b_d$. *Thus, in the slow modulation limit of bath fluctuation, the stochastic perturbation can enhance coherence and give rise to new features.* For all the models when $b_d$ is low we have obtained multiple peaks and sometimes the number of peaks are more than the number of sites present in the model. With increase in $b_d$, the stochastic perturbation vanishes and only the splitting due to the constant intersite coupling (J) survives. Also with increase in the rate of fluctuation each of the peaks becomes sharp and intensity increases. This limit is well-known as the motional narrowing limit, as observed previously.

For the uncorrelated bath case line shapes are broader than that for the correlated bath case. With an increase in the number of site the broadening further increases for uncorrelated bath case because with the increase in the number of sites, the number of uncorrelated bath increases which effectively destroy the mixing of the states. Another important feature is that with increase in the number of site for both linear and cyclic model (line shape of cyclic trimer is provided in the supplementary section) the intensity of the line shape decreases because with increase in number of site the delocalization of exciton increases as well as the coherence increases. The above site dependent study supports the observation of Donehue *et al.* [39] who found that with an increase in ring size, system-bath coupling weakens and intra-molecular coupling increases (in a relative sense).

We also have plotted in the supplementary material the line shape for $J=1$, $V_d=1$, with $b_d$ varying from low value to high value. We have also shown in the supplementary material the line shape in the extreme motional narrowing limit (that is, for very high value of $b_d$ ). We observe that in this limit line shapes are similar for both correlated and uncorrelated baths. However, the difference between correlated and uncorrelated bath increases with increase in the number of sites. Similar results have been obtained earlier for mean square displacement for an initially localized exciton [46].

## C. Effects of off-diagonal fluctuations on line shape

In natural photosynthetic complexes, the off-diagonal fluctuation is negligibly small compare to the diagonal fluctuation. However, the same might not be true for conjugated polymers or other optically active systems. We have thus studied the effect of off-diagonal fluctuation in line shape though we do not provide here any plot of the same. We just mention here that small amount of off-diagonal fluctuation breaks the symmetry of the peaks i.e. shifting of the peak position occurs. However with increase in the off-diagonal fluctuation all the peaks merge with each other. Due to the off-diagonal fluctuation, inter-site coupling $J$ effectively varies. This variation in $J$ is responsible for the mixing of the states with consequent splitting.

In the case of real systems where all the fluctuations are present the line shapes are not symmetric. Scholes *et al.*[35,36] observed two peaks for dimer naphthalene, however the line shape is not symmetric due to the vibrational contribution.

## D. OCCUPATION PROBABILITY FUNCTION (OPF) IN LINEAR AND CYCLIC POLYMERS

In our previous work[50] we have studied occupation probability function (OPF) of the exciton for linear polymer chains of various sizes. Recently Fleming and co-workers,[4,5] Jang *et. al.*[6], Aspuru-Guzik and co-workers[7,8] and Silbey and co-workers[11,12] studied population relaxation dynamics in dimer and FMO complex using different techniques. In this work we have studied and compared the OPF of linear and cyclic tetramer. We have considered correlated and uncorrelated bath cases. We have treated off-diagonal and diagonal fluctuation separately for uncorrelated bath case. For correlated bath case coupled equation of motion can be written as

$$\frac{d\sigma_m}{dt} = -iH_{ex}^x\sigma_m - iV_d\sum_{m'=0}^{1}(\delta_{m+1,m'} + \delta_{m-1,m'})\times\left(\sum_k |k\rangle\langle k|\right)^x \sigma_{m'}$$
$$-iV_{od}\sum_{m'=0}^{1}(\delta_{m+1,m'} + \delta_{m-1,m'})\times\left(\sum_{\substack{k,l\\k\neq l}} |k\rangle\langle l|\right)^x \sigma_{m'} - mb\sigma_m \tag{40}$$

where $O^x f = Of - fO$ . For correlated bath case population of each site is denoted as follows,

$$P_n(t) = \langle n|\sigma_0|n\rangle \tag{41}$$

where $n$ is the site number.

For linear 4 sites model four uncorrelated diagonal fluctuations are available. Hence one can write the coupled equation of motion for this system as follows (we have considered $\hbar = 1$)

$$\frac{d\sigma_{jklm}}{dt} = -iE_0\left(\sum_{k=1}^{4}|k\rangle\langle k|\right)^x \sigma_{jklm} - iJ\left(\sum_{\substack{k=1,l=1\\k\neq l}}^{4}|k\rangle\langle l|\right)^x \sigma_{jklm} - iV_d\sum_{j'=0}^{1}\left(\delta_{j+1,j'} + \delta_{j-1,j'}\right)\left(|1\rangle\langle 1|\right)^x \sigma_{j'klm}$$
$$-iV_d\sum_{k'=0}^{1}\left(\delta_{k+1,k'} + \delta_{k-1,k'}\right)\left(|2\rangle\langle 2|\right)^x \sigma_{jk'lm} - iV_d\sum_{l'=0}^{1}\left(\delta_{l+1,l'} + \delta_{l-1,l'}\right)\left(|3\rangle\langle 3|\right)^x \sigma_{jkl'm}$$
$$-iV_d\sum_{m'=0}^{1}\left(\delta_{m+1,m'} + \delta_{m-1,m'}\right)\left(|4\rangle\langle 4|\right)^x \sigma_{jklm'} - jb_d\sigma_{jklm} - kb_d\sigma_{jklm} - lb_d\sigma_{jklm} - mb_d\sigma_{jklm}. \tag{42}$$

For uncorrelated bath with diagonal fluctuation, population of each site for linear system can be represented as follows,

$$P_n(t) = \langle n|\sigma_{\prod_{N=1}^{N} a_i}|n\rangle \tag{43}$$

where $n$ is the site number, N is the total number of site and $a_1, a_2, a_3, ....a_N$ all the elements are zero. In case of uncorrelated bath with off-diagonal fluctuation population for linear system each site can be represented as follows,

$$P_n(t) = \langle n | \sigma_{\prod_{N=1}^{N-1} a_i} | n \rangle \tag{44}$$

where $n$ is the site number, N is the total number of site and $a_1, a_2, a_3, ....a_{N-1}$ all the elements are zero.

For cyclic 4 sites model three uncorrelated diagonal fluctuations are available ($V_{22} = V_{44}$). Hence one can write the coupled equation of motion for this system as follows (we have considered $\hbar = 1$)

$$\frac{d\sigma_{jkl}}{dt} = -iE_0 \left( \sum_{k=1}^{4} |k\rangle\langle k| \right)^x \sigma_{jkl} - iJ \left( \sum_{\substack{k=1,l=1 \\ k \neq l}}^{4} |k\rangle\langle l| \right)^x \sigma_{jkl} - iV_d \sum_{j'=0}^{1} \left( \delta_{j+1,j'} + \delta_{j-1,j'} \right) \left( |1\rangle\langle 1| \right)^x \sigma_{j'kl}$$

$$-iV_d \sum_{k'=0}^{1} \left( \delta_{k+1,k'} + \delta_{k-1,k'} \right) \left( |2\rangle\langle 2| \right)^x \sigma_{jk'l} - iV_d \sum_{l'=0}^{1} \left( \delta_{l+1,l'} + \delta_{l-1,l'} \right) \left( |3\rangle\langle 3| \right)^x \sigma_{jkl'}$$

$$-jb_d \sigma_{jkl} - kb_d \sigma_{jkl} - lb_d \sigma_{jkl} \tag{45}$$

For uncorrelated bath with diagonal fluctuation, population of each site for cyclic system can be represented as follows,

$$P_n(t) = \langle n | \sigma_{\prod_{N=1}^{N-M} a_i} | n \rangle \tag{46}$$

where $n$ is the site number, N is the total number of site and $M$ arises due to the symmetry criteria (for 3 and 4 sites cyclic model M = 1 and for 6 sites cyclic model M = 2), $a_1, a_2, a_3, ....a_n$ all the elements are zero.

For the cyclic 4 sites model two uncorrelated off-diagonal fluctuations are available ($V_{14} = V_{12}$ and $V_{23} = V_{34}$). Hence one can write the coupled equation of motion for this system as follows (we have considered $\hbar = 1$)

$$\frac{d\sigma_{jk}}{dt} = -iE_0\left(\sum_{k=1}^{4}|k\rangle\langle k|\right)^x \sigma_{jk} - iJ\left(\sum_{\substack{k=1,l=1\\k\neq l}}^{4}|k\rangle\langle l|\right)^x \sigma_{jk} - iV_{od}\sum_{j'=0}^{1}\left(\delta_{j+1,j'}+\delta_{j-1,j'}\right)\left(|1\rangle\langle 2|+|2\rangle\langle 1|\right)^x \sigma_{j'k}$$

$$-iV_{od}\sum_{k'=0}^{1}\left(\delta_{k+1,k'}+\delta_{k-1,k'}\right)\left(|2\rangle\langle 3|+|3\rangle\langle 2|\right)^x \sigma_{jk'} - jb_{od}\sigma_{jk} - kb_{od}\sigma_{jk} \tag{47}$$

For uncorrelated bath with diagonal fluctuation case population for cyclic system each site can be represented as follows,

$$P_n(t) = \langle n|\sigma_{\prod_{N=1}^{N-M} a_i}|n\rangle \tag{48}$$

where $n$ is the site number, N is the total number of site $M$ arises due to the symmetry criteria (for 3 and 4 sites cyclic model M = 1 and 2 respectively and for 6 sites cyclic model M = 3), $a_1, a_2, a_3, \ldots. a_{N-M}$ all the elements are zero.

We have defined normalized occupation probability function as follows,

$$C_n^P(t) = \frac{P_n(t) - P_n(\infty)}{P_n(0) - P_n(\infty)} \tag{49}$$

where $P_n$ is the population of $n^{\text{th}}$ site. We have calculated OPF for uncorrelated bath case with diagonal fluctuation for both 4 sites linear and cyclic model. For uncorrelated bath case with only off-diagonal fluctuation, population for 4 sites linear model can be denoted as $P_n(t) = \langle n|\sigma_{000}|n\rangle$ whereas population for 4 sites cyclic model can be designated as $P_n(t) = \langle n|\sigma_{00}|n\rangle$.

## 1. Comparison of OPF between correlated and uncorrelated bath case (off-diagonal fluctuation)

We have calculated OPF for correlated and uncorrelated bath in slow and fast modulation limit. We have solved coupled equation of motion numerically using Runge-Kutta fourth order method. Dynamics of correlated bath does not involve diagonal fluctuation. Hence to compare the correlated and uncorrelated bath we have calculated OPF for uncorrelated bath using only off-diagonal fluctuation. Change in OPF for uncorrelated bath (diagonal fluctuation) is quite small with increase in rate of fluctuation i.e. going from slow to fast modulation limit because coherence can't be directly destroyed by the diagonal fluctuation.

### A. Slow modulation limit

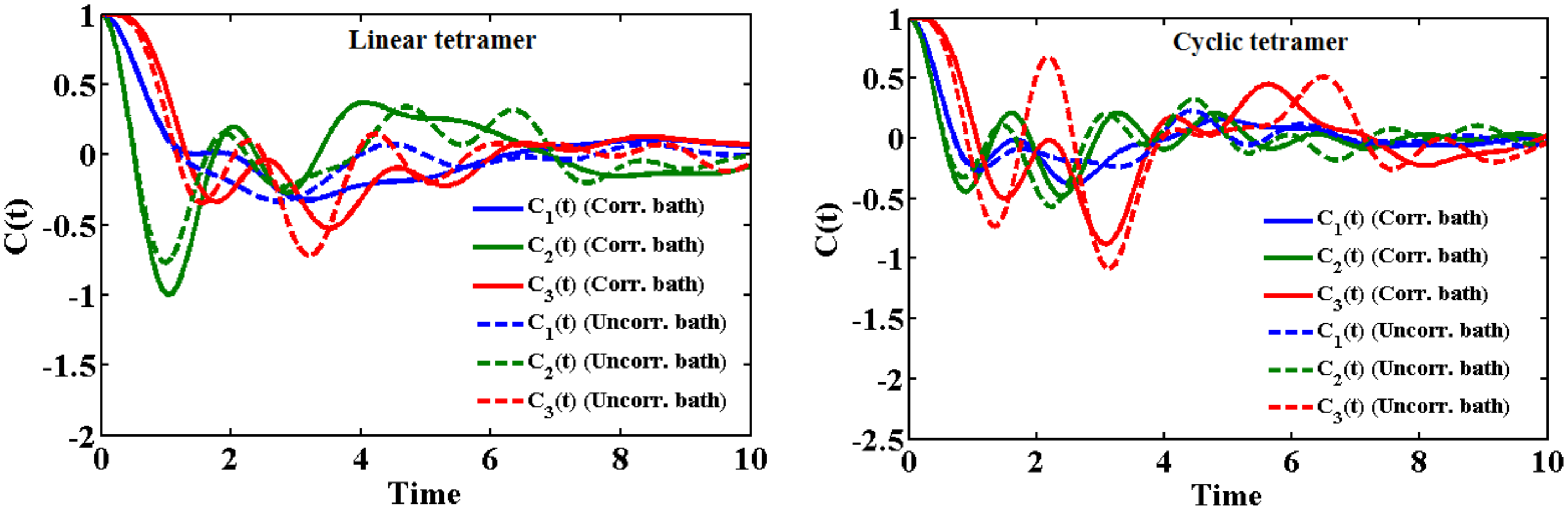

**Fig. 4. Comparison plot of normalized occupation probability functions between the correlated and the uncorrelated bath (off-diagonal fluctuation) in slow modulation limit for (a) linear tetramer (b) cyclic tetramer at J=1, $V_{od}$=0.5 and $b_{od}$=0.5. $C_1(t)$, $C_2(t)$ and $C_3(t)$ are the normalized occupation probability function for site 1, site 2 and site 3 respectively. Solid line corresponds to correlated bath case and dashed line indicates uncorrelated bath case with off-diagonal fluctuation.**

We have compared OPF between the correlated and the uncorrelated bath case (only off-diagonal fluctuation) in slow modulation limit (J=1, $V_{od}$=0.5 and $b_{od}$=0.5) for linear and cyclic tetramer in **Fig. 4**. For both linear and cyclic case correlated and uncorrelated bath OPF shows quite different behavior. Similar observation is obtained from line shape in slow modulation limit for both linear and cyclic models. In slow modulation limit stochastic bath fluctuation increase the coherence and we get large numbers of peaks in line shape. In case of cyclic system interference[58] between two possible path ways increase the amplitude of oscillation in OPF.

## B. Motional narrowing limit

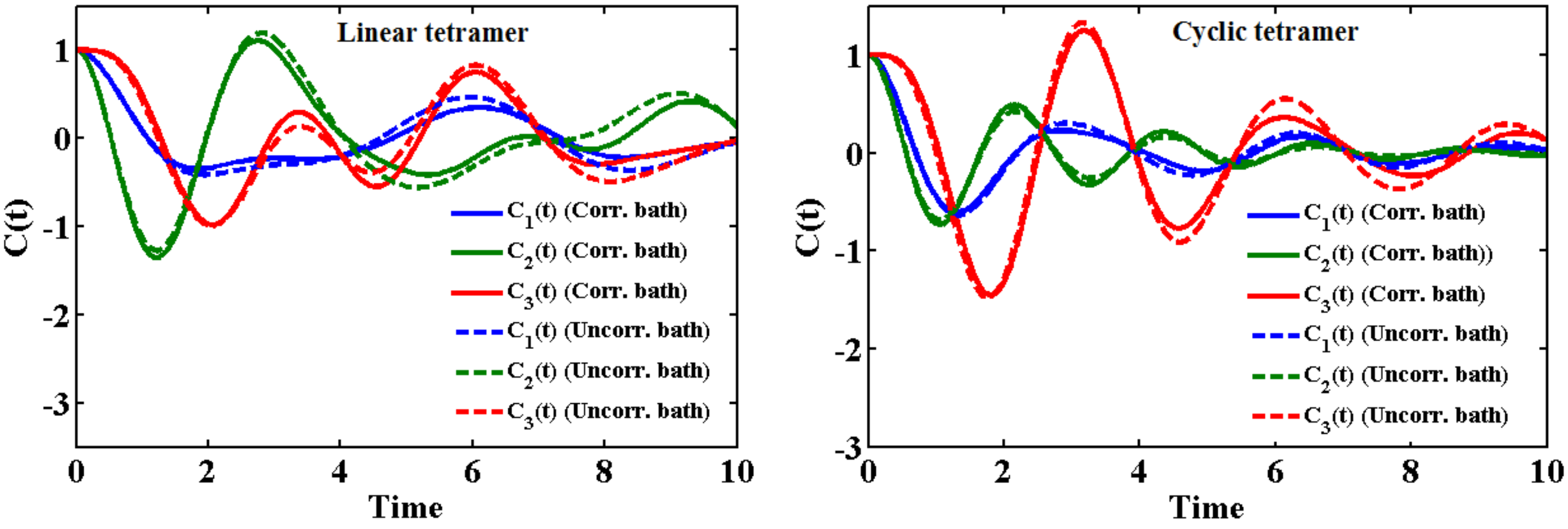

**Fig. 5. Comparison plot of normalized occupation probability functions between the correlated and the uncorrelated bath (off-diagonal fluctuation) in fast modulation limit for (a) linear tetramer (b) cyclic tetramer at J=1, $V_{od}$=0.5 and $b_{od}$=5. $C_1(t)$, $C_2(t)$ and $C_3(t)$ are the normalized occupation**

**probability function for site 1, site 2 and site 3 respectively. Solid line corresponds to correlated bath case and dashed line indicates uncorrelated bath case with off-diagonal fluctuation.**

We have compared OPF between the correlated and the uncorrelated bath case with off-diagonal fluctuation in fast modulation limit for both linear and cyclic tetramer at J=1, $V_{od}$=0.5 and $b_{od}$=0.5 in **Fig. 5**. In fast modulation limit or motional narrowing limit oscillation in OPF for both linear and cyclic model occurs with large amplitude. Both correlated and uncorrelated bath behaves similarly in this limit. Like OPF we also obtain similar line shape for correlated and uncorrelated bath and at high value of rate of fluctuation (extreme motional narrowing limit) the agreement increases. In this limit coherence survives in the system completely due to the inter-site coupling though environment has insignificant contribution to increase the coherence. Also the presence of interference effect between two possible ways of transfer shows sudden increase of oscillatory dynamics in the intermediate time regime for cyclic system.

## E. PROPAGATION OF QUANTUM COHERENCE

Propagation of coherence can be understood by studying the off-diagonal elements present in coupled equation of motion for both correlated and uncorrelated bath case. In this work we have studied off-diagonal elements in equilibrium and excited bath modes for dimer system to examine which one is dominating in slow modulation and fast modulation limit and also to explore the role of decay (either $b_d$ or $b_{od}$) to the propagation of coherence.

### A. Correlated bath case

Fortunately, we can solve the coupled equations of motion of exciton transfer dynamics (equations are provided in the APPENDIX) analytically for dimer system. We assume initially there is no coherence i.e. exciton is initially placed at site 1. We have also solved the population relaxation of each site in both slow modulation and fast modulation limit for larger system. The expression of off-diagonal elements are provided as follows

$$\langle 1|\sigma_0|2\rangle = i\sin 2Jt\exp\left(-\frac{b_{od}}{2}t\right)\left[\frac{b_{od}}{2a}\sinh\frac{a}{2}t+\frac{1}{2}\cosh\frac{a}{2}t\right] \tag{50}$$

$$\langle 1|\sigma_1|2\rangle = \frac{2iV_{od}}{a}\cos 2Jt\exp\left(-\frac{b_{od}}{2}t\right)\sinh\frac{a}{2}t \tag{51}$$

where, $a=\sqrt{b_{od}^2-16V_{od}^2}$

To study the coherence we provide plots of off-diagonal elements i.e. coherence in equilibrium and excited bath state for both slow modulation and fast modulation limit as follows

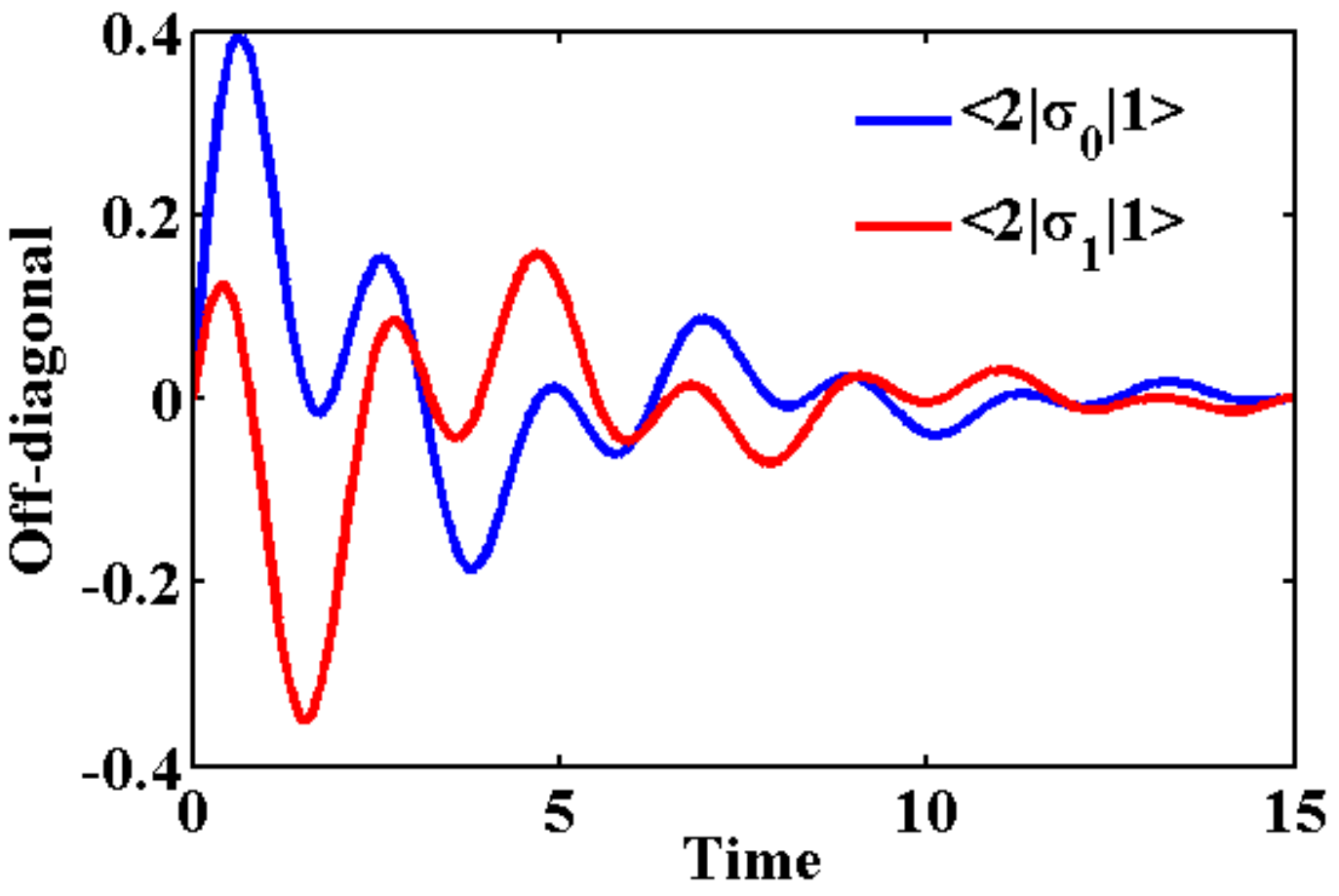

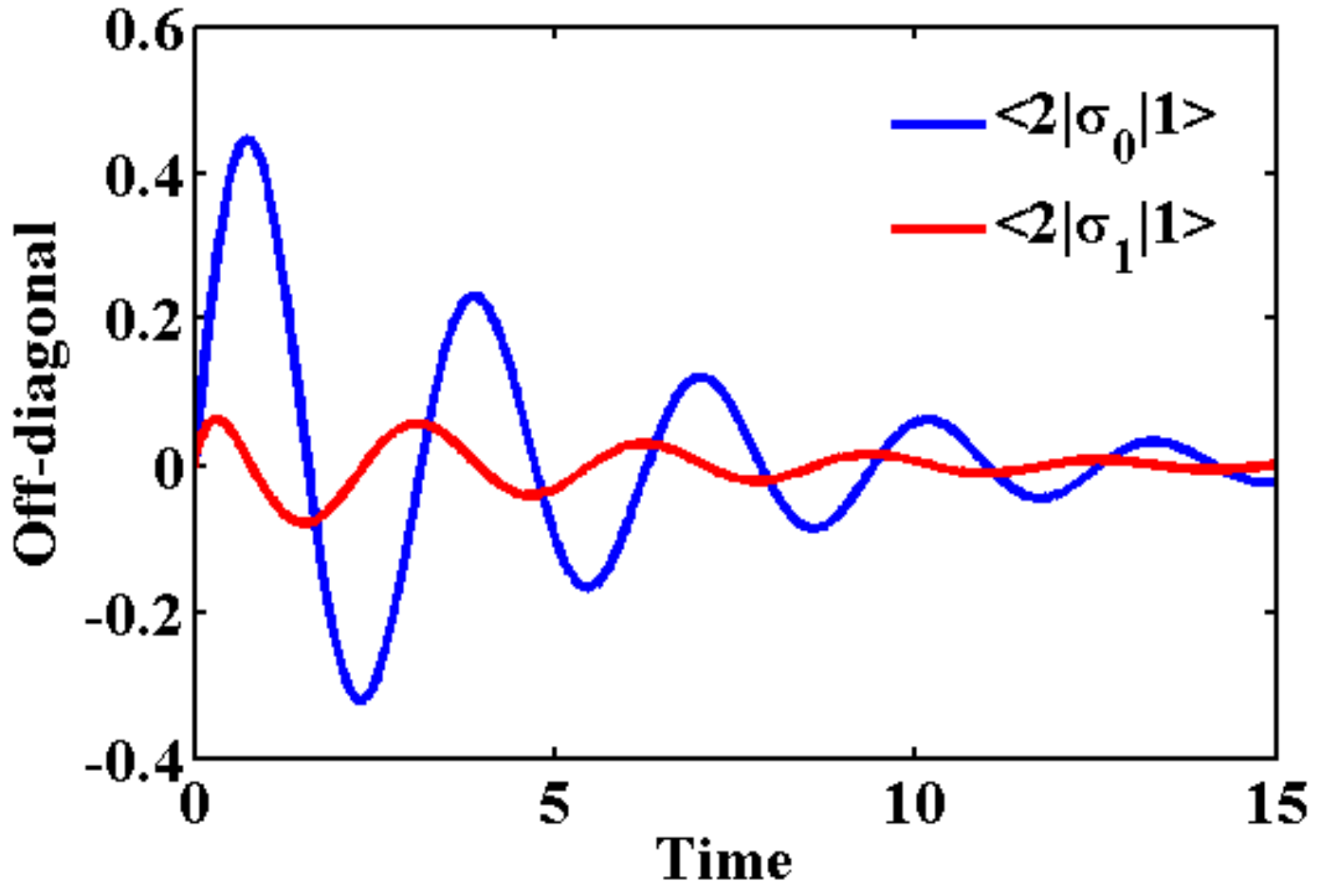

**Fig. 8. Off-diagonal elements are plotted for the dimer problem against time for correlated bath case. (a) Propagation of off-diagonal elements in equilibrium and excited bath modes respectively in Slow modulation limit at J=1, $V_{od}$=0.5 and $b_{od}$=0.5. (b) Propagation of off-diagonal elements in equilibrium and excited bath mode respectively in fast modulation limit at J=1, $V_{od}$=0.5 and $b_{od}$=5.**

All the elements plotted above are imaginary in nature. This can be understood from Eq. (A1) and (A3). As population of each site is real, off-diagonal elements in equilibrium and excited bath modes have to be imaginary. In slow modulation limit described here, both the decay of $\langle 2|\sigma_0|1\rangle$ and $\langle 2|\sigma_1|1\rangle$ occurs with the same rate. In fast modulation limit or motional narrowing limit $\langle 2|\sigma_1|1\rangle$ decays more rapidly than that of $\langle 2|\sigma_0|1\rangle$. Hence one can conclude that for correlated bath case, in slow modulation limit both the decay of equilibrium and excited bath modes occurs in similar time. However, in motional narrowing limit excited bath modes decay rapidly than that of equilibrium bath modes.

## B. Uncorrelated bath case (diagonal fluctuation)

In case uncorrelated bath total number of coupled equation of motion is large and we only provide plots of off-diagonal elements in both slow modulation and fast modulation limit.

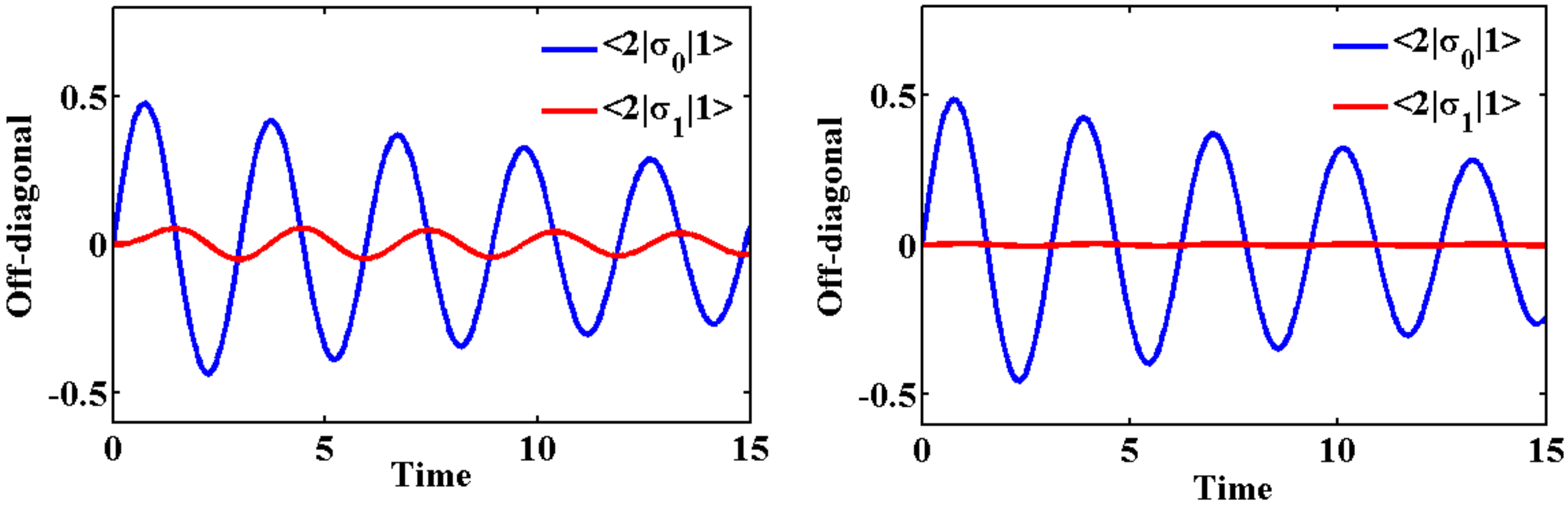

**Fig. 9. Off-diagonal elements are plotted against time for the dimer problem for uncorrelated bath case (off-diagonal fluctuation) (a) imaginary part of $\langle 2|\sigma_{00}|1\rangle$ and imaginary part of $\langle 2|\sigma_{11}|1\rangle$ in slow modulation limit at J=1, $V_{od}$=0.5 and $b_{od}$=0.5. (b) imaginary part of $\langle 2|\sigma_{00}|1\rangle$ and imaginary part of $\langle 2|\sigma_{11}|1\rangle$ in fast modulation limit at J=1, $V_{od}$=0.5 and $b_{od}$=5.**

It interesting to note that unlike the correlated bath case, for uncorrelated bath case in both slow modulation limit and motional narrowing limit, off-diagonal element in equilibrium bath mode dominates over the off-diagonal element in the excited bath modes. In slow modulation limit decay of the off-diagonal element in both bath modes occur in same time. However in motional narrowing limit the decay of the off-diagonal element is rapid that it is flat in nature though the decay time of off-diagonal element in equilibrium bath modes does not change.

Hence from both correlated and uncorrelated bath case one can conclude that in slow modulation limit both off-diagonal element in equilibrium and excited bath modes is responsible for the oscillation of OPF. However, in motional narrowing limit the contribution in oscillatory OPF completely arises due to the off-diagonal element in equilibrium bath modes.

## F. EFFECT OF TEMPERATURE ON LINE SHAPE AND ENERGY TRANSFER DYNAMICS

Stochastic Liouville equation includes the assumption of infinite bath temperature**.** For this reason temperature effect in the line shape and energy transfer dynamics can't be clearly explain using QSLE. Few years ago Tanimura[57] have included the temperature correction terms in QSLE. However, without using QSLE, one can explain qualitatively the effect of temperature on line shape and energy transfer dynamics. Effect of change in temperature will be pretty small in line shape. As electronic energy difference between two levels is quite large, temperature effect is quite small. Change in temperature only can alter the vibrational energy levels i.e. responsible for vibrational relaxation during the electronic excitations. Temperature effect enters through the parameter space $V_d$ and $b_d$ largely while $J$ has weak temperature dependence. With increasing temperature, diagonal fluctuation increases rapidly i.e. why line shape broadens and maximum intensity decreases though the change is quite small.

However, temperature strongly infuences the energy transfer dynamics. At low temperature fluctuations are quite small and i.e. why superposition of states which are responsible for long lived coherent energy transfer dynamics. When temperature is high fluctuations are quite large and superpositions between the states i.e. coherence is destroyed and energy transfer occurs through incoherent hopping mechanism. Low temperature effect can be manifested in the energy transfer dynamics through the large amplitude of oscillation in OPF. However, at high temperature the decay of OPF will be over-damped.

## G. CONCLUSION

The present study investigates the nature of optical line shape in both linear chains and cyclic two level discrete model systems of different sizes, with both diagonal disorder and off-diagonal disorder, in both Markovian and non-Markovian limits. Additionally, we have calculated population transfer dynamics in the same systems in the same limits. This allowed us to interrogate signatures of coherence in line shape.

Among various system sizes investigated, we have focused in this work on a (i) dimer, (ii) a linear chain of tetramer and (iii) a cyclic tetramer. Many of the features observed are familiar from earlier line shape studies, such as a crossover from a static modulation to a motional narrowing limit with increase in the bath relaxation rate. There are, however, several new features arise from the extended nature of the system. Let us summarize the main results of the work.

(1) First and foremost, coherence in the extended systems of the type modeled here arises mostly from the presence of a constant inter-site coupling, denoted here by J. That is, the matrix element J needs to be non-zero. An interesting case arises when J is zero but off-diagonal coupling between sites is slowly varying and large. We discuss this point later.

(2) We have also discussed how one can analyze the nature of the line shape using property of Toeplitz matrix. In the absence of any fluctuations but with constant off-diagonal coupling J, the Hamiltonian of the system is a tridiagonal Toeplitz matrix. Now the eigenvalues of this matrix in our special case is exactly known. Therfore, we know exactly the peak positions of the line shape that we obtain in the fast modulation limit. This identity can be particularly useful because one can fit the peak positions to obtain the value of the off-diagonal coupling parameter J. Actually, this is the method used earlier to obtain the values of J from experimental lineshape of a dimer. But to the best of our knowledge, this has not been implemented for larger chain or ring polymers.

(3) In case of monomer we observe a single sharp Lorentzian peak in the fast modulation limit (large $b_d$). With decrease in $b_d$, the line shape broadens but one does not recover the Gaussian shape as anticipated in Kubo's simple theory. This suggests that Kubo's theory, based on cumulant expansion, may not be effective in the case of Poisson bath and for mixed quantum-classical systems. In the latter case the bath statistics experienced by the system may be significantly non-Gaussian, as discussed long time ago by Oxtoby.[17]

(4) In the extreme slow modulation limit, *the system-bath off-diagonal coupling between different sites can act as a promoter of coherence between the sites and as a result the spacing between the peaks become larger than that given by J*. Interestingly, in limiting cases, the number of peaks can be greater than the total number of sites. In this interesting limit, a less numbers of peaks is observed for correlated bath case than for uncorrelated bath, because presence of large number of uncorrelated baths effectively destroy the mixing between the states. The correlated and uncorrelated baths show profoundly different behavior in this limit. Such effects are particularly important when the average off-diagonal coupling J is very small or zero but the fluctuating element ($V_{od}$) is large.

(5) In the fast modulation limit or the motional narrowing limit, any presence of coherence in the system due to the constant inter-site coupling (J) can be destroyed by fluctuations arising from bath degrees of freedom. Hence splitting of the peaks or number of the peaks for linear system follows the total number of sites and for cyclic system one can observe cyclic symmetry in the number of peaks. In this limit both correlated and uncorrelated bath shows similar line shape.

(6) In the slow modulation limit, correlated and uncorrelated bath cases both exhibit profoundly different behaviors both in OPF and in line shape. In this limit bath plays a crucial role to enhance the coherence and/or increase the oscillatory dynamics.

(7) In the fast modulation limit or motional narrowing limit, the role of bath fluctuation reduces and population transfer dynamics shows less oscillatory behavior. For both linear and cyclic models we observe similar population transfer dynamics for correlated and uncorrelated baths. In this limit both the line shape for correlated and uncorrelated bath behaves similarly. Also the intensity of the peaks are higher than that in the slow modulation case.

(8) We have also discussed the existence of coherence in a dimer system in the presence of both correlated and uncorrelated bath cases. In the slow modulation limit for correlated bath case, we have observed decay of off-diagonal elements in both equilibrium and excited bath modes occurs with the same rate. *However, in the fast modulation limit off-diagonal element in the excited bath modes decays more rapidly than those in equilibrium bath mode*. The situation is somewhat different in the case of uncorrelated bath in slow modulation limit. Here off-diagonal elements in the equilibrium bath mode dominate over the off-diagonal elements in the excited bath mode, although the decay occurs in same time. However, in motional narrowing limit the decay of off-diagonal element in excited bath mode is so rapid that it is essentially zero in the relevant time range where equilibrium mode decay.

(9) A well-known limitation of QSLE is its inability to relax the system to the Boltzmann energy distribution and hence its limitation in describing effects of temperature. Temperature effect in excitonic line shape could be small as the electronic transition energy is much larger than the thermal energy. However, with increasing temperature rate of fluctuation increases rapidly consequently line shape becomes sharp and maximum intensity of all the peaks increases.

We now discuss the relevance of our theoretical study to a few recent experimental observations. First, let us point out that the theory predicts that with increasing number of sites in the system, the intensity of each peak decreases in a certain definite fashion (which is apparent from the Toeplitz matrix and which is proper to compare only within a particular type of model).

This suggests that our theory can be used to explain, at this stage qualitatively, the experimental results of Donehue *et. al.*,[39] who have observed that with increasing the ring size, system-bath coupling weakens and intra-molecular coupling increases due to large delocalization of exciton in synthetic cyclic light harvesting pigment. In our study we have also observed that with increase in ring size intensity decreases. In our model this decrease happens due to the delocalization of the wave function that is the same as observed by Donohue et al. This can be interpreted as an increase of effective coupling between the sites and a decrease of system-bath interaction. In several important studies, first Zewail et al. and subsequently Scholes *et. al.*[35,36] observed splitting of the absorption band for dimer tetrachloro benzene and dimer naphthalene, respectively. This can also be explained from the present theory as we also find for the dimer system in the fast modulation limit appearance of two well-separated peaks. As mentioned above, we can invoke the analytical result via Toeplitz matrix to obtain an estimate of coupling parameter J. We hope to address in a future work a detailed comparison between the theory developed here and the available experimental results.

The present study needs to be generalized in several different directions because models employed here are rather idealistic in the sense that Haken-Strobl-Reineker-Silbey Hamiltonian ignores static disorder that could be prevalent in real systems (like in conjugated polymers). First, it may be worthwhile to study the effects of static or quenched randomness in the energy of the two level systems on both the exciton migration process and the optical lineshape. Second, the effect of static randomness in the off-diagonal coupling J can be quite interesting to investigate. These problems may be studied partly by using random Toeplitz matrix which itself is a problem of much current interest. The static randomness in both diagonal and off-diagonal energies brings this problem close to that of Anderson localization. It will be worthwhile to study effects of dynamic disorder in these problems.

## ACKNOWLEDGEMENT

We thank Dr. Rajib Biswas and Dr. Sarmistha Sarkar for many discussions. We thank the Department of Science and Technology (DST, India) for partial support of this work. B Bagchi thanks Sir J. C. Bose Fellowship for providing partial financial support.

## APPENDIX

Here we provide the coupled equations of motion for correlated bath case given as follows

$$\langle 1|\dot{\sigma}_0|1\rangle = iJ\langle 1|\sigma_0|2\rangle - iJ\langle 2|\sigma_0|1\rangle + iV_{od}\langle 1|\sigma_1|2\rangle - iV_{od}\langle 2|\sigma_1|1\rangle \tag{A1}$$

$$\langle 1|\dot{\sigma}_1|1\rangle = iJ\langle 1|\sigma_1|2\rangle - iJ\langle 2|\sigma_1|1\rangle + iV_{od}\langle 1|\sigma_0|2\rangle - iV_{od}\langle 2|\sigma_0|1\rangle - b\langle 1|\sigma_1|1\rangle \tag{A2}$$

$$\langle 2|\dot{\sigma}_0|2\rangle = -iJ\langle 1|\sigma_0|2\rangle + iJ\langle 2|\sigma_0|1\rangle - iV_{od}\langle 1|\sigma_1|2\rangle + iV_{od}\langle 2|\sigma_1|1\rangle \tag{A3}$$

$$\langle 2|\dot{\sigma}_1|2\rangle = -iJ\langle 1|\sigma_1|2\rangle + iJ\langle 2|\sigma_1|1\rangle - iV_{od}\langle 1|\sigma_0|2\rangle + iV_{od}\langle 2|\sigma_0|1\rangle - b_{od}\langle 2|\sigma_1|2\rangle \tag{A4}$$

$$\langle 1|\dot{\sigma}_0|2\rangle = -iJ\langle 2|\sigma_0|2\rangle + iJ\langle 1|\sigma_0|1\rangle - iV_{od}\langle 2|\sigma_1|2\rangle + iV_{od}\langle 1|\sigma_1|1\rangle \tag{A5}$$

$$\langle 1|\dot{\sigma}_1|2\rangle = -iJ\langle 2|\sigma_1|2\rangle + iJ\langle 1|\sigma_1|1\rangle - iV_{od}\langle 2|\sigma_0|2\rangle + iV_{od}\langle 1|\sigma_0|1\rangle - b_{od}\langle 1|\sigma_1|2\rangle \tag{A6}$$

$$\langle 2|\dot{\sigma}_0|1\rangle = iJ\langle 2|\sigma_0|2\rangle - iJ\langle 1|\sigma_0|1\rangle + iV_{od}\langle 2|\sigma_1|2\rangle - iV_{od}\langle 1|\sigma_1|1\rangle \tag{A7}$$

$$\langle 2|\dot{\sigma}_1|1\rangle = iJ\langle 2|\sigma_1|2\rangle - iJ\langle 1|\sigma_1|1\rangle + iV_{od}\langle 2|\sigma_0|2\rangle - iV_{od}\langle 1|\sigma_0|1\rangle - b_{od}\langle 2|\sigma_1|1\rangle \tag{A8}$$